# Exploiting Dynamic Workload Variation in Low Energy Preemptive Task Scheduling

Lap-Fai Leung, Chi-Ying Tsui[1]
Department of Electrical and Electronic Engineering
Hong Kong University of Science and Technology
Clear Water Bay, Hong Kong SAR, China
{eefai,eetsui}@ee.ust.hk

Xiaobo Sharon Hu[2]
Department of Computer Science and Engineering
University of Notre Dame
Notre Dame, IN 46556, USA
shu@cse.nd.edu

## Abstract

A novel energy reduction strategy to maximally exploit the dynamic workload variation is proposed for the offline voltage scheduling of preemptive systems. The idea is to construct a fully-preemptive schedule that leads to minimum energy consumption when the tasks take on approximately the average execution cycles yet still guarantees no deadline violation during the worst-case scenario. End-time for each sub-instance of the tasks obtained from the schedule is used for the on-line dynamic voltage scaling (DVS) of the tasks. For the tasks that normally require a small number of cycles but occasionally a large number of cycles to complete, such a schedule provides more opportunities for slack utilization and hence results in larger energy saving. The concept is realized by formulating the problem as a Non-Linear Programming (NLP) optimization problem. Experimental results show that, by using the proposed scheme, the total energy consumption at runtime is reduced by as high as 60% for randomly generated task sets when comparing with the static scheduling approach only using worst case workload.

## 1. Introduction

Energy consumption is one of the critical design issues in real-time embedded systems (RTES), which are prevalent in many applications such as automobiles, and consumer electronics, etc. RTES are generally composed of a number of tasks to be executed on one or more embedded processors. Dynamic voltage scaling (DVS), i.e., varying the supply voltage and the corresponding clock frequency of a processor at runtime according to the specific performance constraints and workload, is proven to be very effective for reducing energy consumption [1,2]. Many modern embedded processors support both variable supply voltage and the controlled shutdown mode [3,4]. How to maximally exploit the benefit provided by such hardware has been an active research topic during the last several years. In this paper, we focus on real-time embedded preemptive systems using variable voltage processors.

Having an effective voltage schedule, i.e., the voltage to be used at any given time is critical to harvest the DVS benefit. There are two main approaches to find a voltage schedule. One category of approaches [5] determines the schedule during runtime only. These results can work with either real-time or non-real-time tasks. The basic principle is that only the runtime workload information which is predicted during the online phase is used to determine the voltage schedules. Although such approaches have been shown to result in energy saving, they do not exploit the fact that much information about tasks in an RTES, such as task periods, deadlines, worst-case execution cycles (WCEC) and average workload, is available offline. It is not difficult to see that not using such information may lose opportunities to further reduce the energy consumption.

To complement the above runtime approaches, the other category of voltage scheduling work finds the desired voltage schedules offline based on the available task information, e.g., [1,2,6,7,8,9,10]. These techniques are generally applicable to real-time tasks with hard deadlines. To ensure that the schedule obtained in offline does not violate any timing constraint, the worst-case execution cycles (WCEC) of each task is always used in the offline analysis. Such offline voltage schedules can be altered to some extent at runtime by using the slacks resulted from the tasks not executing at the WCEC to lower the voltage obtained in the offline phase [2,7]. The effectiveness of the offline approach together with the runtime approach is very much dependent on how the slacks are distributed, which in turn depends on the end-time obtained in the static schedule. Therefore, it is important to schedule the tasks in such a way that the potential slack times can be maximally exploited. For many real-time systems, most of the time the workload of the tasks are much smaller than the worst case and on average the execution cycles of the tasks are close to an average-case execution cycle value (ACEC) instead of the WCEC. In general, the schedules obtained from the WCEC values can greatly limit the flexibility and effectiveness of utilizing the slacks generated from the actual execution cycles during runtime.

In this work, a novel offline scheduling approach, which results in the best slack distribution in terms of energy saving for the ACEC scenarios yet guarantees no deadline violation when tasks assume WCEC, is introduced. We focus on preemptive systems, which are more complicated, and it is easily to transform the formulation for non-preemptive systems. To the best of our knowledge, this is the first work that incorporates the ACEC and the WCEC together during the offline variable voltage scheduling. Given that the workload distribution of many real-times can be estimated offline (e.g., using profiling [11]), our approach can achieve much higher energy saving. Experimental results show that significant energy reduction is achieved when the ACEC is considered during the offline scheduling phase.

[1]This work was supported in part by the Hong Kong Research Grant Council under Grant CERG HKUST 62149/03E and HKUST grant HIA02/03.EG03.
[2]This work was supported in part by U.S. National Science Foundation under grant number CCR02-08992 and CNS04-10771.

## 2. Preliminaries and Motivation

### 2.1 System model

In this paper we assume a frame-based preemptive hard real time system in which a frame of length $L$, which is the hyper-period of the task-sets, is executed repeatedly. Rate monotonic (RM) scheduling policy is used to schedule the periodic tasks where the shorter the period of the task, the higher the priority. The priorities of two tasks are the same if they have the same period. A higher priority task will always preempt the current task. We assume no blocking section is available for a task and hence a higher priority task will preempt the lower priority tasks immediately once it is released. The tasks are assumed to be independent of each other. Our technique works for both dependent and independent tasks as well as for multiple processors. For simplicity, we only consider the single processor case in this paper.

Without loss of generality, a set of $N$ periodic tasks is denoted as $\{T_1, T_2, ..., T_N\}$ with $T_i$ has a higher priority than $T_j$ if $i<j$. Each task $T_i$ has its own period $P_i$, the Worst-Case-Execution-Cycles (WCEC) $\hat{W}_i$ and the Average-Case-Execution-Cycles (ACEC) $W_i$. The ACEC is defined as the expected value of the execution cycle base on the workload distribution and it can be obtained by profiling techniques [11]. The relative deadline is assumed to be equal to the period $P_i$. Each task $T_i$ releases its $j^{th}$ instance $T_{i,j}$ periodically. The first instance of all the tasks is assumed to be released at time t=0. Also, each task instance $T_{i,j}$ has its own absolute release time $R_{i,j}$ and absolute deadline $D_{i,j}$. The $P_i$ and the relative deadline of each instance of the task are assumed to be the same. For a lower priority task instance $T_{i,j}$, it may be preempted by others during execution and hence it will be divided into several sub-instances and each sub-instance is denoted as $T_{i,j,k}$ where $k=\{1,..,K\}$ if $T_{i,j}$ is preempted into K sub-parts. When there is no preemption for the task $T_{i,j}$, the task instance itself is denoted as $T_{i,j,1}$ in order to have a consistent notation. Also, we denote the number of task instances of $T_i$ be $Ni_i$ and the upper bound of the number of sub-instances be $NS_{i,j}$.

### 2.2 Motivational example

In this sub-section, we use a non-preemptive system as a motivation example to illustrate the idea of exploiting the workload variation for voltage scheduling. The main idea for preemptive and non-preemptive system is the same except that the formulation of the problem is different. The problem formulation for the preemptive system will be discussed in Section 3.

Let $C_i$ be the effective switching capacitance and $v_i$ be the supply voltage of task $T_i$. The cycle time, $CT$, and the task $T_i$'s execution time $d_i$ can be computed as

$$CT = \frac{\lambda \cdot v_i}{(v_i - V_{th})^\alpha} \quad (1) \quad , \quad d_i = W_i \cdot CT = W_i \cdot \frac{\lambda \cdot v_i}{(v_i - V_{th})^\alpha} \quad (2)$$

where $V_{th}$ is the threshold voltage, $\lambda$ is a device related parameter and $\alpha$ is a process constant which is between 1 and 2. The total energy consumption $e_i$ of executing task $T_i$ is given by

$$e_i = C_i W_i v_i^2 \quad (3)$$

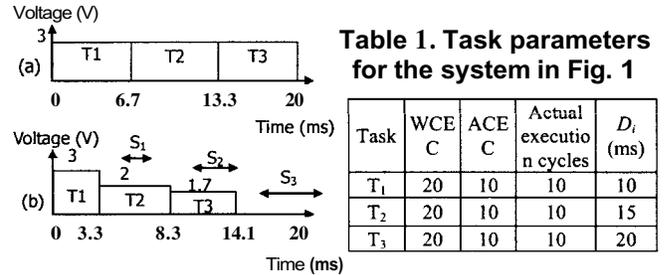

Table 1. Task parameters for the system in Fig. 1

| Task | WCEC | ACEC | Actual execution cycles | $D_i$ (ms) |
|---|---|---|---|---|
| $T_1$ | 20 | 10 | 10 | 10 |
| $T_2$ | 20 | 10 | 10 | 15 |
| $T_3$ | 20 | 10 | 10 | 20 |

**Figure 1. A motivation example**

We use a simple example to illustrate the effect of a static schedule on energy saving when dynamic slack redistribution is employed. Suppose an RTES contains three tasks with the parameters of each task specified in Table 1 (assuming the release time of each task is 0).

Figure 1(a) shows the optimal static schedule if WCEC are taken by all tasks. For simplicity, we assume the clock cycle time is inversely proportional to the supply voltage and the minimum and maximum supply voltages are 0.7V and 5V, respectively. Figure 1(b) gives the actual dynamic run-time schedule when greedy dynamic slack redistribution is carried out. The supply voltage value at runtime depends on both the WCEC and the end-time obtained in the static schedule and can be computed by equation (2). During runtime, tasks finish earlier since their actual execution cycles are smaller than the WCEC. Greedy slack distribution distributes all the slack obtained from the just-finished task to the next task. For example, slack $S_1$ obtained from task $T_1$ is 3.3ms as shown in Figure 1(b) and is utilized fully by the next task $T_2$. The supply voltage of $T_2$ is re-calculated based on the WCEC of $T_2$, that is, $v_2=20/(13.3-3.3)=2$. Similarly, slack $S_2$ generated by task $T_2$ is 5ms and $T_3$ can adopt an even lower voltage. By using equation (3), the overall energy consumption for executing the tasks based on the schedule given in Figure 1(a) is 158.9μJ. It is clear that the dynamic slack redistribution indeed leads to more energy saving. However, if we know that the tasks most probably take the ACEC values during actual execution, can we do better?

Let's examine the static schedule in Figure 1 a little bit closer. In this schedule, each task is associated with a predetermined end time, $te$, e.g., $T_1$'s end time is 6.7ms, $T_2$'s is 13.3ms, etc. These end times are then used in the dynamic slack distribution process to compute a new voltage schedule. The static schedule essentially determines the end time for each task. (Note that this *predetermined* end time can be different from the *actual* end time when a task does not assume the WCEC. Since this predetermined end time is used frequently in our discussion, we simply call it the end time). Such end times are obtained so that the tasks will complete by their deadlines and the overall energy is minimum if tasks take on the WCEC. Now, consider a different schedule where the end times of each task is given as follows: the end times of $T_1$, $T_2$ and $T_3$ are 10, 15 and 20 ms, respectively. Using this schedule and the same greedy slack distribution as above, we obtain the runtime schedule as shown in Figure 2(a). The overall energy consumption of the schedule is $120\mu J$, a 24% improvement comparing with that of the schedule in Figure 1(b).

Though the schedule used by Figure 2 leads to a bigger energy saving, it is important that the schedule can still meet the deadline requirement when tasks assume the WCEC. It is true that the schedule dictates that the end time of task is no later than its deadline. However, if the schedule is not carefully chosen, the tasks may not be able to finish by their deadlines during runtime. Figure 2(b) shows what happens under the schedule used in Figure 2(a) if the tasks do take the WCEC during runtime. At time zero, a 2V is adopted for $T_1$. Since $T_1$ takes the WCEC, it will not finish until $10ms$. The voltages for $T_2$ and $T_3$ can be computed accordingly. Note that 4V is needed for both $T_2$ and $T_3$ in order to meet the timing constraints. If the maximum voltage level for the processor is 3.3V, the schedule would not be feasible. Therefore, simply using the task deadlines as the desired end times does not always give a feasible schedule.

We would like to point out that the actual schedule in Figure 2(b), when tasks happen to take the WCEC, consumes $720\mu J$ energy, a 33% increase over the schedule in Figure 1(a). However, in general, actual execution cycles of a task tend to be close to an average case value and only rarely equal the WCEC value. Based on this observation, we would like to find a static schedule that result in better energy saving on average but still satisfy the timing requirements for the worst case. Even though the above example deals with the non-preemptive schedule only, the basic idea is the same with preemptive scheduling and we will discuss how to formulate the problem of preemptive schedule in the next sections.

In the preemptive system, a task will be preempted into several sub-instances and how to assign the optimal workload for each sub-instance to obtain overall minimum average energy consumption is a challenging problem. With the optimal workload assignment, we can find the corresponding end-time in the static schedule. The static end-time as well as the WCEC for each sub-instances will thus be used for the calculation of the runtime supply voltage.

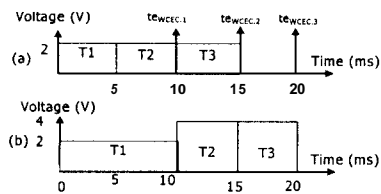

**Figure 2. Another schedule for the system in Fig. 1.**

## 3. Our Approach

From the discussion above, we can see that the greedy slack distribution (or any other slack distribution) relies heavily on the tasks' end time obtained in the static schedule. Existing static voltage scheduling techniques employ the WCEC in order to guarantee that no deadline violation occurs during runtime. Because of the use of the WCEC, the end time of each task is usually more conservative. If we could extend the end time of each task to as long as that allowed by the worst-case execution scenario, it will have more potential for the dynamic slack distribution to achieve more energy saving for the average cases. So our problem is that given the effective switching capacitance, the workload distribution, WCEC, release time and deadline of each task, find a desired schedule, i.e., the desired end time of each task, which strive to maximize the potential energy saving when the tasks are executing based on the workload distribution.

In this section, we show that this scheduling problem can be formulated as a mathematical programming problem. We ignore the voltage transition overhead in our formulation. In most RTES applications, the task execution time is much longer than the voltage transition time. As stated in [12], the increase of energy consumption is negligible when the transition time is small comparing with the task execution time. In the rest of this section, we adopt the following convention: $x$ and $\hat{x}_i$ indicate the average and the worst case values of **x**, respectively. For example, $W_{i,j,k}$ and $\hat{W}_{i,j,k}$ are the average execution cycles and the worst case execution cycles of task sub-instance $T_{i,j,k}$, respectively.

### 3.1 Fully Preemptive Schedule

In our formulation, we want to find the static end-time for each sub-instance by optimally assigning the workload so that the average energy consumption is minimum while all the worst-case requirements are satisfied. In variable voltage scheduling, task's execution time varies inversely with the supply voltage by equation (2). With a longer execution time, the number of preemption by the other higher priority tasks, and hence the number of task sub-instances, is higher because the overlapping region with the higher priority tasks is larger. In order to ensure feasibility of the final schedule and to allow maximum flexibility for the mathematical programming to find the optimal assignment, we need to consider all the possible preemption and the maximum number of task sub-instances. Here we construct a fully preemptive schedule which reflects all possible preemptions based on the periods and priorities of the tasks. All the possible sub-instances of the task instances are found in this schedule. Figures 3 and 4 show an example of how to obtain the fully preemptive schedule. Suppose we have three tasks with $P_1=3$, $P_2=4$ and $P_3=6$. The initial task instances for a hyper-period are shown in Figure 3. All possible preemptions to a task instance are obtained and the original schedule is expanded to a fully preemptive schedule as shown in Figure **4.** However, the actual run-time schedule may not be the same with this schedule because the lower priority task may finish execution before the higher priority released. Here, we want to deal with a more general case that the optimal workload required for each sub-instances are found during the mathematical programming formulation.

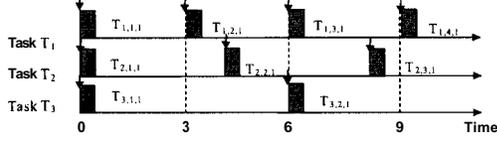

**Figure 3. Task instances in the hyper-period of an example system.**

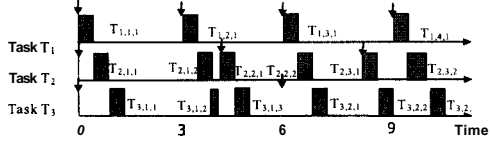

**Figure 4. A fully preemptive schedule for the system in Fig. 3.**

From the fully preemptive schedule, we can obtain the order of the execution of the tasks' sub-instances which is based on the priority and the release time of each sub-instance. E.g. $T_{2,1,2}$ is preempted by $T_{1,2,1}$ and so $Order_{2,1,2} > Order_{1,2,1}$. Since $T_{2,1,1}$ and $T_{2,1,2}$ are originated from the same task instance, we have $Order_{2,1,2} > Order_{2,1,1}$. The total order in Figure **4** is given by:
(1,1,1) <(2,1,1) <(3,1,1)<(1,2,1)<(2,1,2)<(3,1,2)<(2,2,1)<(3,1,3) <(2,2,2)< (3,2,1)< (2,3,1)< (3,2,2)< (1,4,1)< (2,3,2)< (3,2,3).

## 3.2 Problem Formulation

Determining the schedule that optimizes the energy consumption for a preemptive task-set based on the tasks workload distribution while satisfying the timing requirement in the worst case can be formulated as a Non-Linear Programming (NLP) problem. The model consists of an objective function that minimizes the average energy consumption of the system when the tasks take on some workload distribution while subject to a set of resource and timing constraints. The interesting part of this formulation is the way we relate the average execution cycle based on the probability density function of the workload and the worst case execution cycle. If the probability density function is not known, we can use the ACEC as an approximation. In [7], it is shown that this is a good enough approximation of the average energy consumption. We assume the processor can use any voltage value within a specified range. In the following formulation, $T_{i,j,k}$ denotes the current task sub-instance and $T_{i',j',k'}$ is the previous task sub-instance based on the order of the fully preemptive schedule. Next, we define the variables that we are going to find:

$ts_{i,j,k}$    Average start-time of $T_{i,j,k}$
$te_{i,j,k}$    End-time of $T_{i,j,k}$
$w_{i,j,k}$    Average workload of $T_{i,j,k}$
$\hat{w}_{i,j,k}$    Worst-case workload of $T_{i,j,k}$
$v_{i,j,k}$    supply voltage of $T_{i,j,k}$ based on average workload
$\hat{v}_{i,j,k}$    Supply voltage of $T_{i,j,k}$ based on worst-case workload

Among these variables, only the end-time $te_{i,j,k}$, and the worst-case workload variables will be passed to the online DVS phase to calculate the runtime supply voltage. In order to satisfy the worst-case requirements during runtime, the value of $te_{i,j,k}$ and $\hat{w}_{i,j,k}$ will be determined suitably together with the other variables when solving the NLP. It is important to note that the average start-time depends only on the average workload of the previous task but not the average workload of itself since the start time depends on the slack available from the previous tasks. However, the end-times are the same for both the average-case and the worst-case workload conditions. The complete NLP formulation is described as follows.

The objective function of the NLP formulation is

$$\text{Min.} \sum_{i=1}^{N} \sum_{j=1}^{Ni} \sum_{k=1}^{Ns_{i,j}} C_i \cdot w_{i,j,k} \cdot v_{i,j,k}^2 \qquad (4)$$

The probability weighted workload can be used in the objective function if the probability density function is known. Here, we use the average workload in the formulation.
To meet the release time and deadline requirements as well as the voltage range requirement, the following constraints are used:

$$R_{i,j} \leq ts_{i,j,k} \qquad (5)$$

$$te_{i,j,k} \leq D_{i,j} \qquad (6)$$

$$V_{min} \leq v_{i,j,k}, \hat{v}_{i,j,k} \leq V_{max} \qquad (7)$$

$$te_{i,j,k} = ts_{i,j,k} + \hat{w}_{i,j,k} \frac{\lambda \cdot v_{i,j,k}}{(v_{i,j,k} - v_{th})^{\alpha}} \qquad (8)$$

Also, we need to make sure that there is enough allowable working time between the end time of $T_{i',j',k'}$ and $T_{i,j,k}$ for $T_{i,j,k}$ to finish if both tasks use WCEC. We express this by the following constraint:

$$te_{i,j,k} - te_{i',j',k'} \geq \hat{w}_{i,j,k} \frac{\lambda \cdot \hat{v}_{i,j,k}}{(\hat{v}_{i,j,k} - v_{th})^{\alpha}} \qquad (9)$$

If we do not consider the dynamic slack distribution, we would need $te_{i',j',k'} \leq ts_{i,j,k}$ in order to ensure that no task executions are overlapped. Allowing the slacks of finished tasks to be utilized by the subsequent tasks can be thought of as the average start time of $T_{i,j,k}$ becomes earlier than the scheduled end time of $T_{i',j',k'}$ if $T_{i',j',k'}$ uses ACEC instead of WCEC. Assume that the greedy slack distribution is used, the difference between $te_{i',j',k'}$ and $ts_{i,j,k}$ is bounded by the difference of the worst case execution time and the average case execution time, i.e., the slack of $T_{i',j',k'}$. Therefore, we have the following constraint:

$$ts_{i,j,k} \geq te_{i',j',k'} - \frac{\lambda \cdot (\hat{w}_{i',j',k'} - w_{i',j',k'}) \cdot \hat{v}_{i',j',k'}}{(\hat{v}_{i',j',k'} - v_{th})^{\alpha}} \qquad (10)$$

Now, we need to determine the workload assigned for each task sub-instance. The sum of the workloads of the sub-instances is equal to the workload of the task instance because each sub-instance executes only part of the work of its parent task instance. We assume the workload of every instance of the task is the same and hence $W_{i,j} = W_i$ and we have

$$W_i = \sum_{k=1}^{Ns_{i,j}} w_{i,j,k} \qquad (11)$$

$$\hat{W}_i = \sum_{k=1}^{Ns_{i,j}} \hat{w}_{i,j,k} \quad (12)$$

The average workload is always less than or equal to the worst-case workload, so we have:

$$w_{i,j,k} \leq \hat{w}_{i,j,k} \quad (13)$$

From equations (11) and (12), we can see that there are many combinations of $w_{i,j,k}$ and $\hat{w}_{i,j,k}$ with the sums are equal to $W_i$ and $\hat{W}_i$. To find an optimal value for each of them, we divide the workload distribution of all the sub-instances into three cases. Here, we need to explain the meaning of the average workload of the sub-instances. It represents the amount of workload that should be executed on that particular sub-instance when the task instance takes the ACEC. For example, the ACEC and WCEC of a task instance are equal to 15 and 30, respectively. Also, it is preempted into three sub-instances and all of them with WCEC equals to 10. This means that each of them can execute up to 10 units of execution cycles. During the average-case scenario, the first sub-instance will execute 10 units but not 5 units (15/3 units) because the next sub-instance will start execution only if the previous sub-instance already reaches the worst-case limit. With the same argument, the ACEC of the second and third sub-instances are 5 and 0 units, respectively. The ACEC of the third sub-instance is 0 unit means that this sub-instance does not need to perform any execution during the average-case while it is still reserved with enough time slots when the actual execution needs the worst-case cycles. In this case, all the sub-instances need to perform 10 units' execution cycles.

Now we formulate the above idea in the form of mathematical programming. For each sub-instance $T_{i,j,k}$, it falls into one of the following cases: (case 1) $\sum_{k'=1}^{k} \hat{w}_{i,j,k'} < W_i$; (case 2) otherwise. From the above discussion, we can see that to satisfy the average workload distribution, we have $\hat{w}_{i,j,k'} = w_{i,j,k'}$ for all task sub-instances $T_{i,j,k'}$ that belong to case 1. For case 2 because the average workload will be automatically assigned a suitable value according to the constraint (11) and the fact that the average workload of some of the case 1 task sub-instances have already been assigned. Considering the example shown in Figure 5 where $T_{1,1}$ has three sub-instances. The first sub-instance $T_{1,1,1}$ belongs to case 1 because $\hat{w}_{1,1,1} < W_1$ and we have $\hat{w}_{1,1,1} = w_{1,1,1}$. The second and third sub-instances, $T_{1,1,2}$ and $T_{1,1,3}$, belong to case 2 because $\hat{w}_{1,1,1} + \hat{w}_{1,1,2} > W_1$ and $\hat{w}_{1,1,1} + \hat{w}_{1,1,2} + \hat{w}_{1,1,3} > W_1$. Since $\hat{w}_{1,1,1}$ is already assigned and so we have $w_{1,1,2} = W_1 - \hat{w}_{1,1,1}$ because $T_{1,1,2}$ will execute the remaining workload after $T_{1,1,1}$ finish the execution. Now $T_{1,1,1}$ and $T_{1,1,2}$ have already executed all the required average workload, $T_{1,1,3}$ does not need to carry out any computation on average (i.e. $w_{1,1,3}=0$). Note that all the worst-case workload ($\hat{w}_{1,1,1}, \hat{w}_{1,1,2}, \hat{w}_{1,1,3}$) are non-negative number and the sum of them is equal to $\hat{W}_1$.

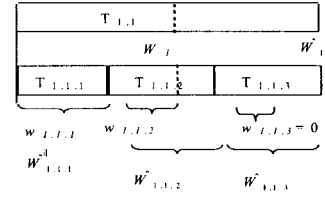

**Figure 5. An example of a task with three sub-instances.**

In the mathematical programming formulation, we deal with a more general case that more than three sub-instances are allowed. However each of the sub-instances still falls into either one of the two cases. The NLP formulation of the above idea is presented as follows.

A dependent linear variable, $ol_{i,j,k} = W_i - \sum_{k'=1}^{k} w_{i,j,k'}$ is introduced to determine whether the current task sub-instance $T_{i,j,k}$ belongs to case 1 or case 2. If it belongs to case 1, i.e. $\sum_{k'=1}^{k} w_{i,j,k'} < W_i$, $ol_{i,j,k}$ is positive. $\sum_{k'=1}^{k} w_{i,j,k'} > W_i$ is impossible because the sum of the average workload of the already executed sub-instances (including the current sub-instance itself) is at most $W_i$. We have $ol_{i,j,k}=0$ if the task sub-instance belongs to case 2. In order to have $w_{i,j,k} = W_i$ for case 1, we have the following additional constraint:

$$w_{i,j,k} \cdot ol_{i,j,k} \geq W_i \cdot ol_{i,j,k} \quad (14)$$

When $\sum_{k'=1}^{k} w_{i,j,k'} < W_i$, i.e. $ol_{i,j,k}>0$, constraint (14) is equivalent to $w_{i,j,k} \geq W_i$. Together with constraint (13), the only feasible solution is $w_{i,j,k} = W_i$. Otherwise, constraint (14) is trivially true when $ol_{i,j,k}=0$.

Finally we need to define the worst-case workload for each of the task sub-instances to yield the best energy saving. However, it is already done since its values are already governed by equations (12) and (13). From the above formulation, solving the NLP problem will results in the optimal assignment of the workload to each sub-instance and the corresponding end-time of each sub-instance will also be obtained.

## 4. Experimental Results

To demonstrate the effectiveness of the proposed technique, which we denote as ACS, a series of experiments, including both randomly-generated task-sets and real-life applications, were carried out. For a given number of tasks, one hundred random task sets were constructed and each taskset results in maximum one thousand of sub-instances. We repeatedly simulated each taskset for one thousand hyper-period. Similar to the experimental settings in [7], we consider the number of execution cycles of each task varying between the best case (BCEC) and worst case (WCEC) following a normal distribution with mean, $\mu = ACEC$, and standard deviation, $\sigma = \dfrac{WCEC - BCEC}{6}$. The BCEC/WCEC ratio is ranging from highly flexible execution (=0.1) to almost fixed (=0.9). The deadline $D_{j,j}$ of each task was chosen from a uniform distribution between 10 and 100. The WCEC of a particular task instance $T_{j,j}$ was adjusted such that the processor utilization is about 70% when all the tasks are running at the maximum speed [7]. We compared the energy consumption using ACS with the energy consumption of the static scheduling method that only considers WCEC in obtaining the scheduling. We denote the later as WCS. The runtime energy consumption is the actual energy consumption after performing the Dynamic Voltage Scaling (DVS) based on either the ACS or WCS static schedules.

Figure 6 summarizes the experimental results. Figure 6(a) shows the comparison between ACS and WCS for different number of tasks when the BCEC/WCEC ratio varies between 0.1(highly flexible execution) and 0.9(almost fixed execution). Y-axis is the percentage improvement in energy consumption of ACS over WCS. It shows that as the number of tasks increases, the energy efficiency of using ACS increases. This can be explained by the fact that as the number of tasks increases, more task sub-instances can use a lower supply voltage by exploiting the workload variation and utilize the slack time generated from the variation. It can be seen that comparing with WCS, the improvement in energy reduction reaches the highest value, about 60% when the BCEC/WCEC value is 0.1 and the number of tasks **is** ten. This is because there are a lot of slacks available when the BCEC/WCEC is low and ACS provides a much better slack utilization in this scenario and minimizes the overall average energy consumption. However, when there is little slack available, i.e., when BCEC/WCEC ratio is high, there is not much improvement as there is little room for both methods to reduce the energy consumption.

To further validate the proposed algorithm, we applied our algorithm to two real-life applications, computer numerical control CNC [13] and GAP [14]. The comparisons of the energy reduction with WCS are shown in Figure 6(b). It can be seen that the improvements over WCS are as high as 41% and 30% when the BCEC/WCEC ratio is 0.1 for CNC and GAP respectively.

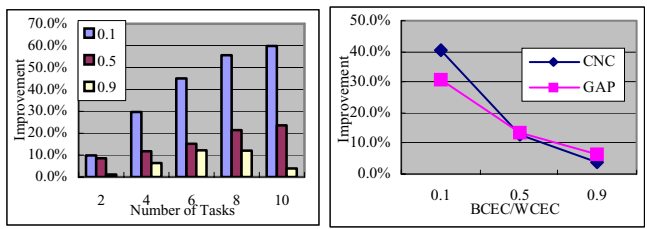

**(a)**      **(b)**
**Figure 6. Experimental results**

## 5. Conclusions

A novel energy reduction strategy in the off-line static voltage scheduling phase was introduced. The preemptive nature of the scheduling **is** considered by using a fully preemptive schedule. The potential slack generated by the later tasks can be utilized by the early tasks by considering the average execution workload during the static voltage scheduling. The problem is formulated as a Non-Linear Programming (NLP) and experimental results showed significant improvement in energy reduction.